\documentclass[11pt,english,twoside]{article}
\pdfoutput=1
\usepackage{jheppub}

\usepackage[english]{babel}
\usepackage{slashed}
\usepackage{amssymb, amsmath, amsthm,latexsym}
\usepackage[english]{babel} 
\usepackage[latin1]{inputenc} 
\usepackage{graphicx}
\usepackage[all]{xy}
\usepackage{bbm}
\usepackage{psfrag}

\newcommand{\be}{\begin{equation}}
\newcommand{\ee}{\end{equation}}
\def\bea#1\eea{\begin{align}#1\end{align}}

\newcommand{\w}{\wedge}

\renewcommand{\i}{\ensuremath{\textnormal{i}}}

\newcommand{\cP}{\mathcal P}

\newcommand{\cN}{\mathcal N}

\newcommand{\beq}{\begin{equation}}
\newcommand{\eeq}{\end{equation}}

\renewcommand{\ee}{\text{e}}

\title{Generalised Geometry and Flux Vacua\\
\small{A very short review}}

\author[a]{Magdalena Larfors\footnote{magdalena.larfors@physics.uu.se}}
\affiliation[a]{Department of Physics and Astronomy, Theoretical Physics, Uppsala university, Box 516, SE-751 20 Uppsala, Sweden}

\abstract{
This note discusses the connection between generalised geometry and flux compactifications of string theory. Firstly, we explain in a pedestrian manner how the supersymmetry constraints of type II $\cN=1$ flux compactifications can be restated as integrability constraints on certain generalised complex structures. This reformulation uses generalised complex geometry, a mathematical framework that geometrizes the B-field. Secondly, we discuss how  exceptional generalised geometry may provide a similar geometrization of the RR fields. Thirdly, we examine  the connection between generalised geometry and non-geometry, and finally we present recent developments where generalised geometry is used to construct explicit examples of flux compactifications to flat space.\\

\vspace{0.5cm}
\noindent
This note is a contribution to the proceedings of "The String Theory Universe" workshop in Leuven, 7-11 Sept 2015, to be published by Fortschritte der Physik.
}
\begin{document}
\maketitle

\section{Supersymmetric Flux Compactifications}

The construction of stable string/M-theory compactifications is a topic that has, for decades, received well-deserved attention from the community of theoretical physicists. One important goal of this endeavour is to construct four-dimensional vacua that model the cosmology and particle physics of  our world. In the most common approach to this problem, one starts by compactifying one of the low-energy supergravity descriptions of M-theory in a way that  preserves $\cN=1$ four-dimensional supersymmetry. To stabilise the extra, compact dimensions, one introduces ingredients such as flux, branes, orientifolds, and instantons. The four-dimensional particle physics is constructed using gauge bundles, intersecting branes, or fields localised at singularities in the compact manifold. Throughout, supersymmetry is used for technical control; supersymmetry breaking is then studied in the effective four-dimensional theory that arises from the compactification.

Already in the early days of heterotic string compactifications, a connection was noted between the $\cN=1$ four-dimensional effective field theory and the geometry and topology of the compact manifold. In compactifications on Calabi--Yau (CY) manifolds, the topology of the internal space determines the field content of the four-dimensional supergravity  \cite{Candelas:1985en}. For instance, the non-trivial Hodge numbers of the CY determine the number of massless scalar fields in four dimensions. It was also noted that supersymmetric vacua can be obtained from non-CY compactifications, if the internal manifold carries a non-zero $H$-flux \cite{Strominger:1986uh,Hull:1986kz}. However, the geometries amenable to such flux compactifications proved more difficult to study than the CY manifolds. Being non-K\"ahler manifolds, they fall outside the class of manifolds mostly studied by geometers, and as a result it is difficult to construct the four-dimensional effective theory resulting from these compactifications. Consequently, non-K\"ahler flux compactifications has received less attention than compactifications on CY manifolds.

However, the many massless scalar fields arising from generic CY compactifcations called for stabilising mechanisms. As noted above, these massless scalars are related to the CY topology, and more specifically to moduli that specify the sizes of the two- and three-cycles of the threefold. Since a $p$-form flux threading a $p$-cycle introduces a potential energy that depends on the size of the cycle, such fluxes may stabilise moduli. Type II supergravity has, in addition to the Neveu--Schwarz (NS) $H$-flux, many $p$-form fluxes in the Ramond--Ramond (RR) sector that may consequently be used to stabilise moduli \cite{Polchinski:1995sm,Michelson:1996pn}. Furthermore, flux compactifications allow to partially break supersymmetry  from $\cN=2$ to $\cN=1$ \cite{Dasgupta:1999ss,Taylor:1999ii,Giddings:2001yu}, and so result in phenomenologically more interesting four-dimensional models. There is by now a huge literature on flux compactifications and the construction of four-dimensional vacua, as reviewed in \cite{Grana:2005jc,Douglas:2006es,Blumenhagen:2006ci,Denef:2008wq}.

Adding flux to a compactification is subtle. As is the case for $\cN=1$ heterotic flux compactifications, the flux in type II compactifications will back react on the compact geometry. Concretely, as we will see in some detail below, the equations of motion and supersymmetry constraints are changed when flux is present, and the geometry of the internal space must therefor adapt, and can no longer remain CY.  Luckily, in contrast to the heterotic case, the constraints thus applied to the geometry can be rewritten in a fashion that generalises the CY constraints, in a precise mathematical sense. Specifically, it can be shown that the manifold must have generalised complex geometry (GCG) \cite{Hitchin:2004ut,Gualtieri04}, as we will now review.\footnote{In compactifications with flux, non-trivial Bianchi identities must be solved, which might be problematic and require the introduction of sources. Furthermore, in compactifications to four-dimensional Minkowski or de Sitter space, the integrated charge and positive stress-energy density associated to the flux must be cancelled by objects with the opposite charge and negative tension \cite{Gibbons:1984kp,Maldacena:2000mw}. In string theory compactifications, orientifold planes provide such sources (see  \cite{McOrist:2012yc} for a discussion of subtleties of type IIA compactifications).}

\section{Generalised Complex Geometry}

For definiteness, we now focus on type IIB string compactifications, noting that the GCG analysis of the type IIA string is completely parallel. Some results carry over to the heterotic string, and we will give some remarks on that in the concluding section. Due to the shortness of this note, this presentation will be brief, and the
reader is referred to the reviews \cite{Grana:2005jc,Zabzine:2006uz,Koerber:2010bx} and references therein for a more  thorough discussion.

The field content of type IIB supergravity comprises of the bosonic NS fields (metric $g$, B-field $B$, dilaton $\phi$), bosonic RR $p$-form potentials ($C_p$ for $p=0,2,4$), and the fermionic fields (gravitini $\psi_M$ and dilatini $\lambda$). A purely bosonic supergravity configuration is supersymmetric if and only if the fermionic supersymmetry variations vanish. As a consequence, any ten-dimensional supersymmetric vacuum must satisfy the Killing spinor equations (KSE) 
\begin{eqnarray}\label{eq:kse}
\delta \psi_M = \left(\triangledown_M + \frac{1}{4} H_M {\cal P} +\frac{1}{16} e^{\phi} \sum_n \slashed{F}_n \Gamma_M {\cal P}_n \right) \epsilon = 0 \; \; , \; \; \\ \nonumber
\delta \lambda= \left( \slashed{\triangledown} \hat{\phi} + \frac{1}{12} \slashed{H}  +
 +\frac{1}{8} e^{\phi} \sum_n (-1)^n (5-n)\slashed{F}_n {\cal P}_n \right) \epsilon= 0 \; .
 \end{eqnarray} 
where $\psi_M$, $\epsilon$ and $\lambda$  are column vectors containing two Majorana--Weyl spinors of the same chirality, $\nabla$~is the standard covariant derivative, and $n$ is odd. Contractions with the ten-dimensional Dirac matrices are denoted $\slashed{\triangledown} = \Gamma^M \triangledown_M$, $\slashed{H} = \Gamma^{MNP} H_{MNP} $, and $H_M = \Gamma^{NP} H_{MNP}$. Finally, the projection matrices $\cP, \cP_n$ are proportional to the Pauli matrices:
\beq 
 \cP=-\sigma^3   \qquad \cP_{3}=\sigma^1
 \qquad \cP_{1,5}=\i\,\sigma^2 
\; .
\eeq
It can be shown that all bosonic equations of motion follow once the KSE and the Bianchi identities are satisfied  \cite{Gauntlett:2005ww,Koerber:2007hd}. Consequently, satisfying the KSE goes a long way  towards constructing supersymmetric string vacua.

To obtain effectively lower-dimensional supersymmetric string vacua, we assume a block diagonal 10-dimensional metric
\beq \label{eq:met}
d s_{10}^2 = e^{2 A(y)} g^{(d)}_{\mu \nu} d x^{\mu} dx^{\nu} + g_{mn} dy^m dy^n\; .
\eeq
We use greek letters to index the coordinates $x^{\mu}$ of the $d$-dimensional non-compact spacetime, and latin indices for the coordinates $y^m$ of the internal manifold that we compactify on. For concreteness we will focus on $d=4$ in the following, but our discussion can be generalised to other values of $d$.  To cohere with the ansatz for the metric, we assume that all supergravity fields decompose accordingly. In particular, the ten-dimensional Killing spinor $\epsilon$ decomposes into four- and six-dimensional spinors, $\zeta_i^a$ and $\eta^i$:
\beq \label{eq:spindec}
\epsilon^A = \sum_{i=1}^{n} \left( \zeta^A_{i+} \otimes \eta^i_+ +
\zeta^A_{i-} \otimes \eta^i_- \right) \; ,
\eeq
where  $\pm$ denotes chiral and anti-chiral components of the spinors,  $\zeta^A_{i-}=\zeta^{A*}_{i+}$, and $\eta^i_-=\eta^{i*}_+$. As a consequence, the four-dimensional supersymmetry depends on the number of supercharges of the ten-dimensional theory (two for type II), and the number of globally defined spinors on the internal space that satisfy the internal part of the KSE \eqref{eq:kse}: one globally defined internal spinor results in $\cN=2$ theories, two such spinors give $\cN=4$ theories. If orientifolds are present, they will partially break supersymmetry.

\subsection{Fluxless Vacua and Complex Structures}
Without flux, the internal part of the \eqref{eq:kse} reduce to the requirement that $\eta^i$ and $\phi=0$ are covariantly constant \cite{Candelas:1985en}. Manifolds that admit a globally defined, covariantly constant spinor have reduced holonomy group. Furthermore, given the existence of a globally defined spinor, one may form globally defined $p$-forms from spinor bilinears. In six dimensions, we have
\beq
J_{mn} = - i \eta^{\dagger}_+ \gamma_{mn} \eta_+ \;  , \;
	\Omega_{mnp} = -i \eta^{\dagger}_- \gamma_{mnp} \eta_+
\eeq
where $\gamma_m$ denote six-dimensional gamma matrices, and the chirality of the spinor is indicated by $\pm$. Let us take a closer look at these forms. $\Omega = \frac{1}{6}\Omega_{mnp} dx^m \w dx^n \w dx^p$ is a complex decomposable three-form, which is closed if $\eta$  is covariantly constant. When this is the case, $\Omega$ defines an integrable complex structure $I$ \cite{Hitchin:2000jd}, with respect to which it is a holomorphic (3,0)-form. 
Similarly, $J = \frac{1}{2}J_{mn} dx^m \w dx^n$ is a real two-form that is closed if $\eta$  is covariantly constant. A closed real two-form on a complex manifold defines a K\"ahler structure. We thus see that $\cN=2$ supersymmetric compactifications of type IIB string theory to four dimensions require the internal 6-fold to be complex, K\"ahler, and have a unique holomorphic top form; the manifold is Calabi--Yau.

To understand how the CY manifolds fit into the larger class of geometries that are generalised complex, it is useful to note a few facts about complex manifolds. The first fact is that an almost complex structure $I$ is a linear map on the tangent bundle of the manifold that squares to -1. As a consequence, it splits the complexified tangent bundle into a holomorphic and an antiholomorphic part corresponding to the $\pm i$-eigenspaces of $I$
\beq
TX \otimes \mathbb{C}= T^{(1,0)} X \oplus T^{(0,1)} X \; .
\eeq
Even-dimensional, orientable manifolds allow such a split locally, but a global decomposition requires that $I$ can be patched consistently across charts. This is possible if the Nijenhuis tensor of $I$ vanishes: in this case the complex structure is integrable, and the ``almost" is dropped when referring to $I$. The existence of a globally defined holomorphic (3,0)-form guarantees that this is the case. Another way of stating this condition is that the Lie bracket of two holomorphic vectors is still holomorphic. In other words, the bundles $T^{(1,0)} X, T^{(0,1)} X$ are involutive with respect to the Lie bracket.

\subsection{Flux Vacua and Generalised Complex Structures}

In the presence of flux, the KSE \eqref{eq:kse}, still require the existence of at least one nowhere-vanishing spinor $\eta$ on the internal manifold, but this is no longer covariantly constant (with respect to the Levi--Civita connection). Hence, we can still define $\Omega$ and $J$ as in the fluxless case but, in the general case, neither form will be closed. Consequently, instead of a CY manifold, we have a manifold with almost complex structure and almost symplectic structure. Using GCG  \cite{Hitchin:2004ut,Gualtieri04}, it can be shown that these manifolds still allow an integrable generalised complex structure.

To show this, we first need to introduce the generalised tangent bundle $E$ of a manifold $X$. Locally, $E$ is $T X \oplus T^* X$, where $T X$ is the tangent bundle, and $T^*X$ the cotangent bundle.\footnote{Globally, $T^* X$ is non-trivially fibered over $TX$, so that $E$ is patched by B-field gauge transformations.\label{fn:gerbe}} Sections of $E$ are called generalised vectors, and are formal sums of vectors and one-forms: $V=v+\xi$, $v \in T X \ , \xi \in  T^*X$. Note that $E$ has twice the dimension of $X$, so for six-manifolds we have a 12-dimensional generalised tangent bundle.

Now, we can discuss generalised complex structures (GCS) on the generalised tangent bundle $E$ in a very similar way to complex structures on the ordinary tangent bundle. A GCS is required to be a linear map ${\cal J}: E \to E$: ${\cal J}^2 = -1_{2d}$. 
Using this, we can decompose $E$ into maximally isotropic (i.e.~$d$-dimensional) subbundles corresponding to the $\pm i$ eigenspaces of ${\cal J}$
\beq
L_{\cal J} = \{ x+ \xi \in E | \frac{1}{2}(1- i {\cal J}) (x+ \xi) = (x + \xi)\} 
\eeq
If $L_{\cal J}$ is involutive with respect to the Courant bracket, which is the generalisation of the Lie bracket to the generalised tangent space, we have an integrable generalised complex structure \cite{Hitchin:2004ut}. 

How do we connect this to $\mathcal{N}=1$ string compactifications? First, there is an analogue of the holomorphic top form $\Omega$ for the generalised complex structure. To define this, we will assume that there are two nowhere vanishing spinors $\eta^{1,2}$ on $X$, noting that we can recover the case with only one such spinor by taking $\eta^1=\eta^2=\eta$. We then form two polyforms, or pure spinors, as follows: 
\beq
\Phi^+ = e^{-\phi} \eta_+^{1} \eta_+^{2 \dagger}  \; , \; 
\Phi^- = e^{-\phi} \eta_+^{1} \eta_-^{2 \dagger} 
\eeq
where $\phi$ is the dilaton. Via the Clifford map, these can be expanded in $p$-forms, and if $\eta^1=\eta^2=\eta$, we have $\Phi^+ = e^{-\phi} e^{i J}$ and $\Phi^- = -i e^{-\phi} \Omega$.\footnote{When $\eta^{1,2}$ are parallel, the structure group of $X$ is SU(3), and when they are perpendicular, it is SU(2). The pure spinor formulation thus allows us to describe these two types of geometries in one go, as well as configurations when the angle between the two spinors vary.} Now, the fact that $\Phi^{\pm}$ are pure spinors on $E$ means that they are annihilated by half of the $\Gamma$ matrices on $E$.  Given a pure spinor $\Phi$, we can then define a six-dimensional subbundle of $E$, called the annihilator space of $\Phi$:
\beq
L_{\Phi} = \{ x+ \xi \in E | (x+ \xi)_A \Gamma^A \Phi = 0\} \; .
\eeq
Consequently, $\Phi$ defines an almost generalised complex structure whose maximally isotropic subspace  $L_{\cal J}$ equals the annihilator space of $\Phi$. It can be shown that $\Phi$ must be closed in order for ${\cal J}$ to be an integrable generalised complex structure. Manifolds with a closed pure spinor are called generalised CY manifolds, using Hitchin's terminology \cite{Hitchin:2004ut}.

Having introduced the pure spinors $\Phi^{\pm}$, we can now establish the connection to ${\cal N}=1$ type II compactifications. In ref.~\cite{Grana:2004bg,Grana:2005sn} it was shown that the KSE \eqref{eq:kse} can be reformulated as the pure spinor equations
\bea \label{eq:ps1}
d (e^{2A} e^{-B} \Phi^+) &= 0 \, \\ \nonumber
d (e^{3A} e^{-B} \mbox{Re} \Phi^-) &= 0 \, \\ \nonumber
d (e^{4A} e^{-B} \mbox{Im} \Phi^-) &= e^{4A} e^{-B} * \lambda(F)  
\eea
where $A$ is the warp factor of the metric \eqref{eq:met}, $B$ is the B-field, and $ \lambda(F) =  F_1-F_3+ F_5 $. Thus, ${\mathcal N}=1$ type IIB vacua are possible on manifolds that allow pure spinors that are conformally closed with respect to the twisted derivative $d_H = d + H \wedge$:
\bea \label{eq:ps2}
d_H (e^{2A} \Phi^+) &= 0 \, \\ \nonumber
d_H (e^{3A} \mbox{Re} \Phi^-) &= 0 \, \\ \nonumber
d_H (e^{4A} \mbox{Im} \Phi^-) &= e^{4A} * \lambda(F)  \; .
\eea
Note that $d_H$ is a nilpotent operator, $d_H^2=0$, owing to the fact that $H$ is closed. This follows from one of the Bianchi identities in type II compactifications. That $d_H$ is nilpotent means that it can be used just as the ordinary exterior derivative $d$, for example to define cohomology groups.  As a consequence, when the RR fluxes are zero, we have a purely algebraic geometrical description of the vacua, and may, just as for CY manifolds, relate infinitesimal deformations of the geometry to cohomology groups. This provides a  model-independent tool to the study of the moduli space of the vacua, which is very valuable in the study of the stability of flux vacua. We will return to this point in the concluding section.

\subsection{Exceptional Generalised Geometry}

As we saw in equations \eqref{eq:ps1}-\eqref{eq:ps2}, the RR flux appear as defect terms in the pure spinor equations, that prevent $\Phi^-$ from being closed, and hence give an integrable generalised complex structure. It is natural to ask if we can define an even more generalised tangent bundle, in which the RR fluxes are treated on par with the geometry, just as the $H$-flux is in GCG. If this is the case, we may hope to find integrable structures that allow an algebraic geometrical analysis of the moduli space of generic $\cN=1$ type II vacua. There are indeed proposals for such geometrizations that go under the name exceptional generalised geometry (EGG) \cite{h07,pw08,go11}, and here we will summarise some important points of these works.

The first thing to note is that the geometrization of the $H$-flux in GCG is tied to the fact that not only diffeomorphisms, but also  B-field gauge transformations, are needed to construct the transition functions of the generalised tangent bundle $E$ (see footnote \ref{fn:gerbe}). Analogously, we expect the generalised tangent bundle that geometrizes both NS and RR fluxes to be patched by diffeomorphisms, and the gauge transformations of $B$ and $C_p$.
Furthermore, as we will discuss in more detail in the next section, T-duality is also integrated in the structure group of $E$.  To geometrize the RR fluxes in a similar manner, we need a bundle whose structure group contains the full U-duality group of type II vacua with NS and RR flux, which in four dimensions is the exceptional group $E_{7(7)}$. 

A so-called exceptional tangent bundle, with the features just described, can be formed. For example, in ref.~\cite{go11} such a  bundle has local decomposition $T X \oplus T^* X \oplus \Lambda^5 T^* X \oplus (T^* X \otimes \Lambda^6 T^*X) \oplus \Lambda^{even} T^* X$, so that its sections are formal sums of vectors, one-forms, five-forms, one-forms tensored by six-forms, and certain polyforms. The twisted derivative of GCG $d_H = e^{-B} d e^{B}$ is generalised to a covariant derivative $D$ twisted by the B-field, its six-dimensional dual and the RR-field polyform $C=C_0+C_2+C_4$: 
\beq
e^B e^{-*_6 B} e^{-C}\; D \; e^C e^{*_6B} e^B \; .
\eeq
Finally, it is shown that the KSE \eqref{eq:kse} do require that a structure on the exceptional tangent bundle is twisted closed, but only upon projection to a certain representation of the U-duality group. Thus, some work remains before a completely algebraic geometrical reformulation of type II $\cN=1$ flux vacua is achieved.

\section{GCG, Non-geometry and F-theory}

As mentioned several times, GCG geometrizes the $H$-flux, or equivalently the B-field. A consequence of this is that T-duality is encoded as part of the structure group of the generalised tangent bundle. In more detail, $E$ has a canonical metric defined by the contraction of vectors with one-forms: 
\beq
h=\begin{pmatrix} 0 & 1 \\ 1 & 0 \end{pmatrix} \ ,\qquad 
\frac{1}{2} \begin{pmatrix} \xi \\ v \end{pmatrix}^T \begin{pmatrix} 0 & 1 \\ 1 & 0 \end{pmatrix} \begin{pmatrix} \xi \\ v \end{pmatrix} = \xi_m v^m \; .
\eeq
It follows that the structure group of $E$ is reduced from $GL(12)$ to $O(6,6)$, since  $O^T h O = h \iff O \in O(6,6)$. $O(6,6)$ includes diffeomorphisms and gauge transformations of $B$, as well as T-duality, which, when described by the Buscher rules \cite{b87,b88}, exchanges components of $g$ and $B$. In addition to the canonical metric $h$, there is one more metric on $E$:  using the pure spinors $(e^{-B} \Phi^+, e^{-B} \Phi^-)$ we can define what is known as the  generalised metric
\beq \label{eq:genmet}
 \mathcal{H}^{MN}=
\begin{pmatrix} g_{ij}-B_{ik} g^{kl}B_{lj} & B_{ik} g^{kj} \\ -g^{ik} B_{kj} & g^{ij}\end{pmatrix}  \; .
\eeq
The generalised metric transforms covariantly under $O(6,6)$ transformations.

The fact that the structure group of the generalised tangent bundle contains T-duality transformations suggests that the GCG formalism may be useful to describe not only $\cN=1$ flux vacua, but also the so-called non-geometric string compactifications. More generally, it may describe any global string compactification composed of local solutions to the KSE and BI that are glued together using $O(6,6)$ (or U-duality) transformations.

\subsection{Non-geometric flux vacua}
A prime example of non-geometric flux vacua are the T-folds, i.e.~configurations patched together using T-duality \cite{dh05,h04}. These configurations are non-geometric, as the required transition functions go beyond the geometric transition functions (diffeomorphisms and gauge transformations), and problematic as supergravity solutions (since sub-string scale cycles are precent), but may still make sense in string theory, where T-dual configurations are equivalent. Moreover, such non-geometric configurations may be T-dual to more standard geometric string compactifications, and therefor seem difficult to exclude when discussing generic compactifications. In GCG, it has been suggested that non-geometry is associated to a certain bi-vector, called $\beta$ \cite{gs06,gs07,Grana:2008yw}. In particular, it was shown by  \cite{Grana:2008yw}, that the generalised metric $\mathcal{H}$ may be parametrized by $g$ and $B$ as in \eqref{eq:genmet}, or by a new metric $\tilde{g}$ and $\beta$:
\beq \label{eq:genmet2}
 \mathcal{H}^{MN}=
\begin{pmatrix} \tilde{g}_{ij} & -\tilde{g}_{ik}\beta^{kj} \\ \beta^{ik} \tilde{g}_{kj} & \tilde{g}^{ij}-\beta^{ik} \tilde{g}_{kl} \beta^{lj} \end{pmatrix} 
\eeq
The two parametrizations correspond to different choices of generalised vielbeine for $\mathcal{H}$.\footnote{We remark that $\tilde{g}$, $\beta$ and $\mathcal{H}^{MN}$ have been used for the discussion of T-duality before the construction of GCG \cite{Duff:1989tf,Duff:1990hn}.}

More recently, this has been used to argue that GCG in the $\beta$-reparametrization provides a ten-dimensional formulation of certain four-dimensional supergravity vacua with non-geometric gaugings. Such a connection was previously difficult to find, as a procedure for dimensional reduction on spaces patched by T-duality was lacking. Ref.~\cite{Andriot:2011uh} proposed that the field redefinition imposed by equation \eqref{eq:genmet} and \eqref{eq:genmet2} in ten dimensions allows to isolate the global non-geometric effects of the action in a boundary term. Discarding this term, a procedure for dimensional reduction can be defined that produces the desired four-dimensional non-geometric gaugings. These ideas have since been developed further, with investigations of formal aspects of the construction as well as studies of examples where the procedure works 
(see \cite{Andriot:2014qla} for recent work and further references).
However, in parallel no-go theorems have been derived, that show that there exist non-geometric configurations that cannot be analysed in this manner  \cite{Dibitetto:2012rk,Lee:2015xga}. Such configurations require a Double Field Theory \cite{Hull:2009mi} description that breaks what is known as the strong constraint of this theory; as such they certainly do not fit into the generalised geometry description. It is also unclear if generalised geometry can be used to describe the non-geometric heterotic vacua that have recently been constructed using heterotic/F-theory duality \cite{McOrist:2010jw,Malmendier:2014uka}. Presently, it seems that generalised geometry does not suffice to give a full understanding of non-geometric string compactifications.

\subsection{F-theory vacua with flux}
 
As already alluded to, we may construct string compactification by solving the the KSE \eqref{eq:kse} on local patches of the manifold, and then glue the solutions together using string dualities. One example of such solutions is in fact F-theory compactifications \cite{Vafa:1996xn}, which are non-perturbative $\cN=1$ compactifications of type IIB string theory, that are patched using S-duality. 

Recently, it was shown that by following a strategy inspired by F-theory, one may solve a recurrent problem in the construction of $\cN=1$ type II Minkowski vacua. In these scenarios, the sources required by the flux backreaction lead to delta-like terms in the supergravity equations, that, as a consequence, can seldom be solved analytically. In $\cN=1$ type IIB compactifications, the warp factor $A$ is governed by a harmonic equation in the compact space that can typically be solved only in the region of weak string coupling \cite{Giddings:2001yu}. However, by gluing together local solutions to the KSE \eqref{eq:kse} in a U-duality consistent manner, one can construct vacua where all supergravity equations are explicitly solved in the presence of fluxes \cite{Martucci:2012jk,Braun:2013yla,Candelas:2014jma,Candelas:2014kma}.

The strategy these references uses relies on the fact that all supergravity fields in certain type flux IIB backgrounds may be mapped to the complex structure moduli of an auxiliary algebraic K3 surface. This is similar to F-theory, where the axio-dilaton field of type IIB supergravity is mapped to the complex structure of an auxiliary elliptic curve. Moreover, just as the modular group of the elliptic curve equals the $SL(2,\mathbb{Z})$ S-duality group of type IIB string theory, the U-duality group of the relevant type IIB backgrounds is encoded in the modular group of the K3 surface. 

In more detail, the backgrounds studied are non-perturbative compactifications of type IIB on SU(2) structure six-folds, composed of  a four-manifold $M^4$ is fibered over a two-sphere. All supergravity fields ($M^4$ metric, B-field, RR potentials $C_p$ and axio-dilaton) are assumed to vary holomorphically over the two-sphere. When $M^4$ equals $T^4$ or K3, it can be shown that these fields will take values in a coset that is isomorphic to the complex structure moduli space of an auxiliary K3 surface. Thus, similar to standard F-theory vacua, we may build flux solutions by studying K3 fibrations.  Since there is an elaborate mathematical machinery for the study of K3 fibrations (including degenerations to singular fibres), this gives an unprecedented method of constructing explicit flux vacua.

As discussed in detail in \cite{Martucci:2012jk,Candelas:2014kma},
the GCG reformulations of the KSE \eqref{eq:ps1} are very useful in  decomposing the supergravity fields into holomorphic functions that can be mapped to complex structure moduli of K3 surfaces. However, the geometrization of the fluxes is very different from that in GCG and EGG. In particular, the latter does not require holomorphicity of the supergravity fields, which is crucial for this treatment of the flux in these generalised F-theory compactifications. It would certainly be relevant to study the relation between these different geometric treatments of flux further. 

\section{Conclusions and Outlook}

In this note, we have discussed the connection between string compactifications with flux and generalised geometry. We started with a short review of generalised complex geometry (GCG), which is the mathematical framework that underlies $\cN=1$ type II compactifications. We noted how the B-field, and the associated $H$-flux, is encoded in GCG in a pure spinor or a twisted exterior derivative. Moreover, we saw how a generalised tangent bundle $E$ can be constructed, and how the structure group of  $E$ includes diffeomorphisms, $B$-field gauge transformations and T-duality transformations. We remarked that a similar geometrization of the RR fluxes of type II theories gives the so-called exceptional generalised geometry (EGG).

We then discussed more recent use of generalised geometry for the description of string compactifications that go beyond the supergravity approximation. We discussed the applicability of generalised geometry to non-geometric string compactifications, as well as its use in constructing explicit examples of F-theory compactifications with flux.

There are many questions that remain to be studied in this field. To conclude, let us mention two aspects that are currently under study. The first is the derivation of a generalised geometry for the heterotic string. 
In GCG, it is important that $H$ is closed. We have mentioned one consequence of this fact, namely that the twisted derivative $d_H$ is nilpotent, but it is used in several important mathematical consistency checks of GCG. In heterotic compactifications, the anomaly cancellation condition implies that $H$ is, in the general case, no longer closed. A generalised geometry for the heterotic string must therefor be formulated in a way that does not require closure $H$. Some steps in this direction have been taken, see 
\cite{Garcia-Fernandez:2013gja,Baraglia:2013wua,Coimbra:2014qaa}.

Second, recall that a major motivation for the inclusion of flux in string compactifications is that they may give masses to the scalar fields associated to geometric moduli. That such a stabilisation occurs  can be shown for the complex structure moduli of warped type IIB CY compactifications with flux, in the large volume limit \cite{Giddings:2001yu}. However, this proof does not hold for generic flux compactification, where the backreacted internal geometry is no longer conformally CY, and a limit where the flux goes to zero might be lacking. Hence, a moduli analysis that does not use the CY approximation is called for, and might take as its starting point the integrability conditions of GCG or EGG, see e.g.~\cite{Martucci:2009sf,Tseng:2011gv}. Recent work in this area also includes studies of the moduli of heterotic and M-theory compactifications, such as \cite{Anderson:2014xha,delaOssa:2014cia,delaOssa:2014lma,Grana:2014vxa}.

\providecommand{\WileyBibTextsc}{}
\let\textsc\WileyBibTextsc
\providecommand{\othercit}{}
\providecommand{\jr}[1]{#1}
\providecommand{\etal}{~et~al.}

\end{document}